\documentclass[12pt]{article}

\usepackage{amsmath}
\usepackage{amssymb,scalefnt}
\usepackage{amsfonts}
\usepackage{latexsym}
\usepackage{color}

\usepackage{ dsfont }

\catcode `\@=11 \@addtoreset{equation}{section}

\catcode `\@=12

\newcommand{\be}{\begin{equation}}
\newcommand{\en}{\end{equation}}
\newcommand{\bea}{\begin{eqnarray}}
\newcommand{\ena}{\end{eqnarray}}
\newcommand{\beano}{\begin{eqnarray*}}
\newcommand{\enano}{\end{eqnarray*}}
\newcommand{\bee}{\begin{enumerate}}
\newcommand{\ene}{\end{enumerate}}

\newcommand{\mc}{\mathcal}

\newcommand{\D}{{\mc D}}

\newcommand{\Sc}{{\cal S}}
\newcommand{\E}{{\cal E}}
\newcommand{\F}{{\cal F}}
\newcommand{\G}{{\cal G}}

\newcommand{\Lc}{{\cal L}}

\newcommand{\1}{1 \!\! 1}

\newcommand{\Hil}{\mc H}

\newtheorem{thm}{Theorem}

\newtheorem{defn}[thm]{Definition}

\catcode `\@=11 \@addtoreset{equation}{section}
\catcode `\@=12

\begin{document}

\thispagestyle{empty}

\vspace*{1.5cm}

\begin{center}
{\Large \bf $\D$-deformed harmonic oscillators}   \vspace{2cm}\\

{\large F. Bagarello}\\
  Dipartimento di Energia, Ingegneria dell'Informazione e Modelli Matematici,\\
Scuola Politecnica Ingegneria, Universit\`a di Palermo,\\ I-90128  Palermo, Italy\\
and I.N.F.N., Sezione di Torino\\
e-mail: fabio.bagarello@unipa.it\\
home page: www.unipa.it/fabio.bagarello

\vspace{2mm}

{\large F. Gargano}\\
Institute for Coastal Marine Environment (IAMC),\\
National Research Council (CNR),\\ I-91026 Mazara del Vallo (TP), Italy\\
e-mail: francesco.gargano@unipa.it\\

\vspace{2mm}

{\large D. Volpe}\\
  Dipartimento di Fisica e Chimica,\\
Universit\`a di Palermo,\\ I-90128  Palermo, Italy\\
e-mail: danielevolpe92@gmail.com\\

\end{center}

\vspace*{.5cm}

\begin{abstract}
\noindent We analyze systematically several deformations arising from two-dimensional harmonic oscillators which can be
described in terms of $\D$-pseudo bosons. They all give rise to exactly solvable models, described by non self-adjoint hamiltonians
whose eigenvalues and eigenvectors can be found adopting the quite general framework of the so-called $\D$-pseudo bosons. In particular, we show that several models previously
introduced in the literature perfectly fit into this scheme.

\end{abstract}

\vspace{2cm}


\vfill


\newpage

\section{Introduction}

In recent years many physicists and some mathematician started to be interested in the possibility of giving a physical meaning to some non self-adjoint hamiltonian with real eigenvalues. The whole story started essentially with Bender and Boettcher in 1998, \cite{ben1}, with the famous $p^2+i\,x^3$ hamiltonian. Since then, hundreds of papers have been written, mainly from  physical and numerical points of view. On the other hand, the mathematically oriented papers were much less. More recently, the number of such papers increased significantly also because several authors, mainly coming from functional analysis and operator algebras, joined the community starting to be interested into this topic. Some of these papers are listed in \cite{mathpap}, \cite{mathpap2} and \cite{mathpap3}, where more references can be found.

In several contributions different authors discuss some aspects of manifestly non self-adjoint hamiltonians which are quadratic in the position and in the momentum operators, or in some of their combinations. Quite often, in their analysis, they are able to deduce the explicit form of the eigenvectors, which are not mutually orthogonal, and of the related eigenvalues, which are real. In some previous papers by one of us (FB), it has been shown that some of the proposed models can be discussed in terms of the so-called $\D$-pseudo bosons ($\D$-PBs), that is of some {\em excitations} arising from properly deformed commutation rules. In this paper we set up a systematic analysis of several two-dimensional models which can be described completely in terms of $\D$-PBs, some of them already considered in the past and others, in our knowledge, new. Doing so, we propose a list of models which are simply {\em $\D$-deformed two-dimensional harmonic oscillators}, for which the approach discussed in, say, \cite{bagnewpb} and reviewed in \cite{bagbook} can be adopted (see also \cite{baggar} for models involving anti-commutation relations).

This article is organized as follows: to keep the paper self-contained, in the next section we review the definition and few results on $\D$ pseudo-bosons ($\D$-PBs). In Section III we discuss in details two models, deducing the eigenvectors of two different non self-adjoint hamiltonians. Among other things, we prove that the eigenstates do not form bases, but they are complete in $\Lc^2({\Bbb R}^2)$. In Section IV we give a list of other hamiltonians which can be discussed using the same techniques, and leading to similar conclusions. Section V contains our final remarks.

\section{$\D$ pseudo-bosons}\label{sectII}

We briefly review here few facts and definitions on $\D$-PBs. More details can be found in \cite{bagnewpb} and \cite{bagbook}. To simplify the notation, we consider here the one-dimensional case, since nothing essential changes going from one to more dimensions\footnote{This is true for commutative models, which are the ones considered all along this paper. When dealing with non-commutative quantum mechanics, differences may arise.}.

Let $\Hil$ be a given Hilbert space with scalar product $\left<.,.\right>$ and related norm $\|.\|$. Let further $a$ and $b$ be two operators
on $\Hil$, with domains $D(a)$ and $D(b)$ respectively, $a^\dagger$ and $b^\dagger$ their adjoint, and let $\D$ be a dense subspace of $\Hil$
such that $a^\sharp\D\subseteq\D$ and $b^\sharp\D\subseteq\D$, where $x^\sharp$ is $x$ or $x^\dagger$. Incidentally, it may be worth noticing
that we are not requiring here that $\D$ coincides with, e.g. $D(a)$ or $D(b)$. Nevertheless, for obvious reasons, $\D\subseteq D(a^\sharp)$
and $\D\subseteq D(b^\sharp)$.

\begin{defn}\label{def21}
The operators $(a,b)$ are $\D$-pseudo bosonic ($\D$-pb) if, for all $f\in\D$, we have
\be
a\,b\,f-b\,a\,f=f.
\label{21}\en
\end{defn}
 Sometimes, to simplify the notation, instead of (\ref{21}) we will simply write $[a,b]=\1$, having in mind that both sides of this equation
have to act on some $f\in\D$.

\vspace{2mm}

Our  working assumptions are the following:

\vspace{2mm}

{\bf Assumption $\D$-pb 1.--}  there exists a non-zero $\varphi_{ 0}\in\D$ such that $a\,\varphi_{ 0}=0$.

\vspace{1mm}

{\bf Assumption $\D$-pb 2.--}  there exists a non-zero $\Psi_{ 0}\in\D$ such that $b^\dagger\,\Psi_{ 0}=0$.

\vspace{2mm}

Then, if $(a,b)$ satisfy Definition \ref{def21}, it is obvious that $\varphi_0\in D^\infty(b):=\cap_{k\geq0}D(b^k)$ and that $\Psi_0\in D^\infty(a^\dagger)$, so
that the vectors \be \varphi_n:=\frac{1}{\sqrt{n!}}\,b^n\varphi_0,\qquad \Psi_n:=\frac{1}{\sqrt{n!}}\,{a^\dagger}^n\Psi_0, \label{22}\en
$n\geq0$, can be defined and they all belong to $\D$. We introduce the sets $\F_\Psi=\{\Psi_{ n}, \,n\geq0\}$ and
$\F_\varphi=\{\varphi_{ n}, \,n\geq0\}$. Since each
$\varphi_n$ and each $\Psi_n$ belong to $\D$, they also belong to the domains of $a^\sharp$, $b^\sharp$, $N=ba$ and $N^\dagger=a^\dagger b^\dagger$. We have
\be
\left\{
    \begin{array}{ll}
b\,\varphi_n=\sqrt{n+1}\varphi_{n+1}, \qquad\qquad\quad\,\, n\geq 0,\\
a\,\varphi_0=0,\quad a\varphi_n=\sqrt{n}\,\varphi_{n-1}, \qquad\,\, n\geq 1,\\
a^\dagger\Psi_n=\sqrt{n+1}\Psi_{n+1}, \qquad\qquad\quad\, n\geq 0,\\
b^\dagger\Psi_0=0,\quad b^\dagger\Psi_n=\sqrt{n}\,\Psi_{n-1}, \qquad n\geq 1,\\
       \end{array}
        \right.
\label{23}\en as well as the following eigenvalue equations: $N\varphi_n=n\varphi_n$ and  $N^\dagger\Psi_n=n\Psi_n$, $n\geq0$. Then,  choosing the normalization of $\varphi_0$ and $\Psi_0$ in such a way $\left<\varphi_0,\Psi_0\right>=1$, we deduce that
\be \left<\varphi_n,\Psi_m\right>=\delta_{n,m}, \label{34}\en
 for all $n, m\geq0$, so that $\F_\varphi$ and $\F_\Psi$ are biorthogonal sets. The third assumption introduced in \cite{bagnewpb} is the following:

\vspace{2mm}

{\bf Assumption $\D$-pb 3.--}  $\F_\varphi$ is a basis for $\Hil$.

\vspace{1mm}

This is equivalent to the request that $\F_\Psi$ is a basis for $\Hil$ as well, \cite{bagnewpb}. In particular, if $\F_\varphi$ and $\F_\Psi$ are Riesz basis for $\Hil$, the $\D$-PBs are called {\em regular}\footnote{We recall that a set $\E=\{e_n\in\Hil,\,n\geq0\}$ is a (Schauder) basis for $\Hil$ if any vector $f\in\Hil$ can be written, uniquely, as an (in general)
infinite linear combination of the $e_n$'s: $f=\sum_{n=0}^\infty c_n(f)e_n$. Here $c_n(f)$ are complex numbers depending on the vector $f$ we
want to expand. If $\E$ is an orthonormal basis, that is when we also have $\left<e_n,e_m\right>=\delta_{n,m}$, then $c_n(f)=\left<e_n,f\right>$.
A  Riesz basis $\F=\{f_n\in\Hil,\,n\geq0\}$ is a set of vectors for which a bounded
operator $T$ on $\Hil$ exists, with bounded inverse, and an orthonormal basis $\E=\{e_n\in\Hil,\,n\geq0\}$, such that $f_n=Te_n$, for all
$n\geq0$. Also, a set of vectors $\F$ is complete if the only vector $h\in\Hil$ which is orthogonal to all the vectors in $\F$ is the zero vector.}

\vspace{2mm}

In \cite{bagnewpb} a weaker version of Assumption $\D$-pb 3 has also been introduced, particularly useful for physical applications: for that, let $\G$ be a suitable dense subspace of $\Hil$. Two biorthogonal sets $\F_\eta=\{\eta_n\in\G,\,n\geq0\}$ and $\F_\Phi=\{\Phi_n\in\G,\,n\geq0\}$ have been called {\em $\G$-quasi bases} if, for all $f, g\in \G$, the following holds:
\be
\left<f,g\right>=\sum_{n\geq0}\left<f,\eta_n\right>\left<\Phi_n,g\right>=\sum_{n\geq0}\left<f,\Phi_n\right>\left<\eta_n,g\right>.
\label{25}
\en
Is is clear that, while Assumption $\D$-pb 3 implies (\ref{25}), the reverse is false. However, if $\F_\eta$ and $\F_\Phi$ satisfy (\ref{25}), we still have some (weak) form of resolution of the identity.  Now Assumption $\D$-pb 3 may be replaced by the following:

\vspace{2mm}

{\bf Assumption $\D$-pbw 3.--}  $\F_\varphi$ and $\F_\Psi$ are $\G$-quasi bases, for some subspace $\G$ dense in $\Hil$.

\vspace{2mm}

Let us now assume that  Assumption $\D$-pb 1,  $\D$-pb 2, and  $\D$-pbw 3 are satisfied. Sometimes, even if it is not strictly necessary, it is convenient to consider $\G=\D$. Then, let us consider a self-adjoint, invertible, operator $\Theta$, which leaves, together with $\Theta^{-1}$, $\D$ invariant: $\Theta\D\subseteq\D$, $\Theta^{-1}\D\subseteq\D$. Hence,  \cite{bagnewpb}, we say that $(a,b^\dagger)$ are $\Theta-$conjugate if $af=\Theta^{-1}b^\dagger\,\Theta\,f$, for all $f\in\D$. We recall that $(a,b^\dagger)$ are $\Theta-$conjugate if and only if  $(b,a^\dagger)$ are $\Theta-$conjugate. Moreover,
 $(a,b^\dagger)$ are $\Theta-$conjugate if and only if $\Psi_n=\Theta\varphi_n$, for all $n\geq0$. Furthermore,
if $(a,b^\dagger)$ are $\Theta-$conjugate, then $\left<f,\Theta f\right>>0$ for all non zero $f\in \D$. These results are all proved in the first paper in \cite{mathpap2}, where also other details of $\D$-PBs, including some interesting intertwining relations, are discussed. In view of the concrete applications considered here it might be useful to stress that, in some explicit models, \cite{bagbook}, $\D$ and $\G$ must be taken different. Of course, this has nothing to do with the Assumptions $\D$-pb 1, $\D$-pb 2, and $\D$-pb 3 (or $\D$-pbw 3).

\section{The two-dimensional deformed harmonic oscillators: two detailed examples}\label{SectIII}
The main ingredient of our construction is a standard two dimensional harmonic oscillator whose hamiltonian can be written as
\be
H_0=\omega_1A_1^\dagger A_1+\omega_2A_2^\dagger A_2+\omega_3\1,
\label{31}
\en
where $\omega_j\in{\Bbb R}$, and $[A_j,A_k^\dagger]=\delta_{j,k}\,\1$, $j,k=1,2$, all the other commutators being zero. It is clear that  $H_0=H_0^\dagger$. The Hilbert space of the model is $\Hil=\Lc^2({\Bbb R}^2)$, with scalar product $\left<f,g\right>=\int_{\Bbb R}\int_{\Bbb R}\overline{f(x_1,x_2)}\,g(x_1,x_2)\,dx_1\,dx_2$.

 What is interesting for us is to consider some {\em deformed} versions of $H_0$, which we write (formally, for the moment) as
$$
H=e^XH_0e^{-X},
$$
for some suitable operator $X$. Because of $(\ref{31})$, introducing (again formally at this stage) $a_j=e^XA_je^{-X}$ and $b_j=e^XA_j^\dagger e^{-X}$, we can write $H$ as follows
\be
H=\omega_1b_1 a_1+\omega_2b_2 a_2+\omega_3\1,
\label{32}
\en
and we have $[a_j,b_k]=\delta_{j,k}\,\1$, $j,k=1,2$. Hence these operators appear to satisfy the pseudo-bosonic commutation rules, since, in general, $b_j\neq a_j^\dagger$, due to the fact that $X^\dagger$ is not assumed to coincide with $-X$.
But, as we have already pointed out several times, this is formal. The reason is simple: most of the times the operators involved in this procedure are unbounded, so that one should pay attention to domain problems. In particular, since $H_0$ is unbounded, if $X$ is also unbounded, there is no reason, a priori, for $H$ to be well defined on a dense (or, even, simply non empty) subset of $\Hil$. This depends, of course, on the form of $X$. For the same reason, the commutator $[a_j,b_k]$ could not be densely defined. This has been discussed, for instance, in \cite{mathpap3}.

In what follows, we will make this transformation concrete and rigorous, by checking in details, for two different choices of $X$,
the validity of Assumptions $\D$-pb 1, $\D$-pb 2 and $\D$-pb 3, or its weaker form, $\D$-pbw 3.
In order to help the reader to identify the expression of $H$ we adopt in this paper, as dynamical variables, the position and momentum operators, rather than raising and lowering operators\footnote{The relation between them is the usual one: $A_j=\frac{x_j+ip_j}{\sqrt{2}}$, $j=1,2$.}. This, we believe, can be useful, although not necessary,  since many models, in the existing  physical literature on the subject, are written adopting this  choice. Therefore we rewrite \eqref{31}  as
\be
H_0=\tilde\omega_1(x_1^2+p_1^2)+\tilde\omega_2(x_2^2+p_2^2)+\tilde\omega_3\1,
\label{33}
\en
 where, to simplify the notation, we have introduced $\tilde\omega_1=\omega_1/2$, $\tilde\omega_2=\omega_2/2$, $\tilde\omega_3=\omega_3-(\omega_1+\omega_2)/2 $. Here, as in ordinary (i.e. commutative) quantum mechanics, $[x_j,p_k]=i\delta_{j,k}\,\1$, $j,k=1,2$, while all the other commutators are assumed to be trivial.

\subsection{The first model}\label{secIII1}

In this section we consider the following quadratic choice for $X$:

\be
X=\gamma(\sqrt{2}(x_1+x_2)+2x_1x_2),
\label{310}
\en
where $\gamma\in {\Bbb R}$. We see that the momentum operators do not appear in $X$, and for this reason $U(\pm\gamma):=e^{\pm X}$ are just (unbounded, and therefore not everywhere defined) multiplication operators. In term of the bosonic operators $A_j$, $A_j^\dagger$, $X$ looks as
\be
X=\gamma(A_1^\dagger A_2+A_2^\dagger A_1+A_1 A_2+A_1^\dagger A_2^\dagger+A_1+A_1^\dagger +A_2+A_2^\dagger),
\label{310b}
\en
which appears rather more complicated.

The operators $a_j$ and $b_j$ introduced before can be rewritten as
\be
\left\{
    \begin{array}{ll}
a_1=(x_1+ip_1)/\sqrt{2}-\gamma\left(x_2\sqrt{2}+\1\right), \quad a_2=(x_2+ip_2)/\sqrt{2}-\gamma\left(x_1\sqrt{2}+\1\right),\\
b_1=(x_1-ip_1)/\sqrt{2}+\gamma\left(x_2\sqrt{2}+\1\right), \quad b_2=(x_2-ip_2)/\sqrt{2}+\gamma\left(x_1\sqrt{2}+\1\right),\\
       \end{array}
        \right.
\label{311}\en
or as
\be
\left\{
    \begin{array}{ll}
a_1=A_1-\gamma\left(A_2+A_2^\dagger+\1\right), \quad a_2=A_2-\gamma\left(A_1+A_1^\dagger+\1\right),\\
b_1=A_1^\dagger+\gamma\left(A_2+A_2^\dagger+\1\right), \quad b_2=A_2^\dagger+\gamma\left(A_1+A_1^\dagger+\1\right).\\
       \end{array}
        \right.
\label{311bis}\en
It is easy to check that they obey, formally, the pseudo-bosonic rule $[a_j,b_k]=\delta_{j,k}\1$, $j,k=1,2$,
the other commutators being zero. However, to go from formal to rigorous results, we now take a completely
different point of view, showing that the four operators \underline{defined} as in (\ref{311}), or as in (\ref{311bis}), satisfy the two-dimensional version of Assumptions $\D$-pb 1, $\D$-pb 2 and $\D$-pbw 3 of Section II. We will also show that assumption $\D$-pb 3 does not hold, so that the vectors we will construct extending (\ref{22}) are not bases for $\Hil$.

First, using (\ref{311}),  we rewrite equations $a_1\varphi_{0,0}=a_2\varphi_{0,0}=0$ in a differential form as
$$
\frac{\partial\varphi_{0,0}(x_1,x_2)}{\partial x_1}+\left(x_1-\gamma(2x_2+\sqrt{2})\right)\varphi_{0,0}(x_1,x_2)=0
$$
and
$$
\frac{\partial\varphi_{0,0}(x_1,x_2)}{\partial x_2}+\left(x_2-\gamma(2x_1+\sqrt{2})\right)\varphi_{0,0}(x_1,x_2)=0.
$$
The solution is easily deduced:
\be
\varphi_{0,0}(x_1,x_2)=N\,e^{-\frac{1}{2}(x_1^2+x_2^2)+\sqrt{2}\,\gamma(x_1+x_2)+2\gamma x_1 x_2},
\label{312}\en
where $N$ is a normalization constant we will fix in the following. Now, since $b_j^\dagger$ coincides with $a_j$, but with $\gamma$ replaced by $-\gamma$, it is clear that the solution of  $b_1^\dagger\Psi_{0,0}=b_2^\dagger\Psi_{0,0}=0$ can be deduced from (\ref{312}) with a similar replacement:
\be
\Psi_{0,0}(x_1,x_2)=N'\,e^{-\frac{1}{2}(x_1^2+x_2^2)-\sqrt{2}\,\gamma(x_1+x_2)-2\gamma x_1 x_2}.
\label{313}\en
Here $N'$ is another normalization constant, which needs not to coincide with $N$. A direct computation shows that, if $\gamma\notin\left]-\frac{1}{2},\frac{1}{2}\right[$, then both $\varphi_{0,0}(x_1,x_2)$ and $\Psi_{0,0}(x_1,x_2)$ are not square integrable.  On the other hand, if we take $\gamma\in\left]-\frac{1}{2},\frac{1}{2}\right[ $ and we choose $N=N'=\frac{1}{\sqrt{\pi}}$, then $\left<\varphi_{0,0},\Psi_{0,0}\right>=1$. Among other things this means that, even if formally $(a_j,b_j)$ satisfy the pseudo-bosonic rules for all possible values of $\gamma$, they are surely not $\D$-pseudo bosonic operators if $\gamma\notin\left]-\frac{1}{2},\frac{1}{2}\right[$. For this reason, from now on we will assume that $\gamma \in \left]-\frac{1}{2},\frac{1}{2}\right[ $. In this case both $\varphi_{0,0}(x_1,x_2)$ and $\Psi_{0,0}(x_1,x_2)$ are square integrable and, more than this, they belong to $\Sc({\Bbb R}^2)$,  the set of $C^\infty$-functions which decrease to zero, together with their derivatives, faster than any inverse power. For future convenience, it is also worth noticing that they also both belong to the set
$$
\D:=\left\{f(x_1,x_2)\in\Sc({\Bbb R}^2):\,  e^{\beta_1x_1+\beta_2x_2}f(x_1,x_2)\in\Sc({\Bbb R}^2),\, \forall \beta_1,\beta_2\in{\Bbb C}\right\}.
$$
This set, which we are taking as our $\D$, is dense in $\Lc^2(\Bbb R^2)$, since it contains $D(\Bbb R^2)$, the set of the $C^\infty$ functions with compact support. Moreover, as required, $\D$ is stable under the action of both $a_j^\sharp$ and $b_j^\sharp$, and in fact we deduce that, by using a two-dimensional version of (\ref{22}),
\be
\varphi_{n_1,n_2}(x_1,x_2)=\frac{\tilde N}{\sqrt{2^{n_1+n_2}\,n_1!\,n_2!}}\,H_{n_1}(x_1)H_{n_2}(x_2)\,e^{-\frac{1}{2}(x_1^2+x_2^2)+\sqrt{2}\gamma(x_1+x_2)+2\gamma x_1x_2},
\label{313b}\en
for all $n_j\geq0$. Here $\tilde N$ is a suitable normalization, related to $N$, which is not particularly important in our analysis. Notice that, not surprisingly, each $\varphi_{n_1,n_2}$ belongs to $\D$ (so that, in particular, it belongs to $\Sc(\Bbb R^2)$). Incidentally, it is worth mentioning that $\D$ is invariant also under the action of $e^{\pm X}$.

The function $\Psi_{n_1,n_2}(x_1,x_2)$ can be deduced by $\varphi_{n_1,n_2}(x_1,x_2)$ simply replacing $\gamma$ with $-\gamma$. Therefore, also $\Psi_{n_1,n_2}(x_1,x_2)$ belong to $\D$. Our conclusion, so far, is that Assumptions $\D$-pb 1 and $\D$-pb 2 are indeed satisfied. To check whether $\D$-pb 3 is also satisfied, we first define the sets $\F_\Psi=\{\Psi_{n_1,n_2}(x_1,x_2),\,n_j\geq0\}$ and $\F_\varphi=\{\varphi_{n_1,n_2}(x_1,x_2),\,n_j\geq0\}$. Their vectors are eigenstates of $N_j$ and $N_j^\dagger$: $N_j\varphi_{n_1,n_2}=n_j\varphi_{n_1,n_2}$ and $N_j^\dagger\Psi_{n_1,n_2}=n_j\Psi_{n_1,n_2}$, $j=1,2$, and are mutually orthogonal. To show that neither $\F_\Psi$ nor $\F_\varphi$ are bases we use a standard argument, \cite{dav}: we prove that the norms of $\varphi_{n_1,n_2}(x_1,x_2)$ and $\Psi_{n_1,n_2}(x_1,x_2)$ are both divergent for diverging $n_1$ or $
 n_2$.

As a matter of fact, due to the relation between these vectors, it is enough to check that $\|\varphi_{n_1,n_2}\|$ diverges with $n_j$. Moreover, it is sufficient to consider the case when $n_2=0$, and to prove that $\|\varphi_{n,0}\|\rightarrow\infty$ when $n\rightarrow\infty$. Indeed, since $H_0(x)=1$, we easily find that
$$
I_n:=\frac{\|\varphi_{n,0}\|^2}{\tilde N^2}=\frac{1}{2^n\,n!}\sqrt{\frac{\pi}{1-4\gamma^2}}\,e^{\frac{4\gamma^2}{1-2\gamma}}\int_{\Bbb R} H_n\left(\frac{t+t_0}{\sqrt{1-4\gamma^2}}\right)^2\,e^{-t^2}\,dt.
$$
The integral can be now estimated by first changing variable: let us put $s=\frac{t}{\sqrt{1-4\gamma^2}}$. Then, since
$$
\int_{\Bbb R} H_n\left(\frac{t+t_0}{\sqrt{1-4\gamma^2}}\right)^2\,e^{-t^2}\,dt \geq \sqrt{1-4\gamma^2} \int_{\Bbb R} H_n(s+s_0)^2\,e^{-s^2}\,ds,
$$
where $s_0=\frac{t_0}{\sqrt{1-4\gamma^2}}=\frac{\sqrt{2}\,\gamma}{1-2\gamma}$, we deduce that
$$
I_n\geq\pi \,e^{\frac{4\gamma^2}{1-2\gamma}}L_n\left(-\left(\frac{2\gamma}{1-2\gamma}\right)^2\right),
$$
which, for all non zero allowed $\gamma$, diverges when $n\rightarrow\infty$, see \cite{szego}. A similar estimate can be repeated for $\|\Psi_{n,0}\|^2$. Hence, $\F_\varphi$ and $\F_\Psi$ are not bases. Still, it is possible to check that they are both complete in $\Lc^2(\Bbb R^2)$. This follows, for instance, from their analytical expression, see (\ref{313b}), and by a simple extension of the completeness argument for functions of the form $x^n f(x)$, where $|f(x)|\leq Ce^{-\delta x}$, $\delta>0$, and $n=0,1,2,\ldots$, to a two-dimensional case, see \cite{kolm}, pg 426\footnote{The same conclusion can be deduced following \cite{sjo}, Lemma 3.12}.
Now, even if $\F_\varphi$ and $\F_\Psi$ are not bases, Assumption $\D$-pbw 3 could still be true. This is important since, as it is proved in \cite{mathpap2,bagnewpb}, this milder condition is enough to deduce several interesting consequences.

We first observe that, since $a_j\varphi_{0,0}=U(\gamma)A_jU^{-1}(\gamma)\varphi_{0,0}=0$, then $\varphi_{0,0}$ belongs to the domain of  $U^{-1}(\gamma)=U(-\gamma)$. Moreover, because of the uniqueness of the vacuum of $A_j$, $\Phi_{0,0}(x_1,x_2)$, we deduce that $\varphi_{0,0}(x_1,x_2)$ must be proportional to $U(\gamma)\Phi_{0,0}(x_1,x_2)$. Similarly, $\Psi_{0,0}$ belongs to the domain of  $U^\dagger(\gamma)=U(\gamma)$, and it must be proportional to $U(-\gamma)\Phi_{0,0}(x_1,x_2)$. Moreover, because of condition $\left<\varphi_{0,0},\Psi_{0,0}\right>=1$, we can fix these proportionality constants as follows:
\be
\left\{
    \begin{array}{ll}
\varphi_{0,0}(x_1,x_2)=U(\gamma)\Phi_{0,0}(x_1,x_2),\\
\Psi_{0,0}(x_1,x_2)=U(-\gamma)\Phi_{0,0}(x_1,x_2).\\
\end{array}
        \right.
        \label{314}
\en
Incidentally, this is in agreement with our previous remark on the role of $\gamma$ in $\varphi_{0,0}(x_1,x_2)$ and $\Psi_{0,0}(x_1,x_2)$. Now we can check explicitly that: (i) $\varphi_{n_1,n_2}(x_1,x_2)\in D(U(-\gamma))$, for all $n_j\geq0$; (ii) $\Psi_{n_1,n_2}(x_1,x_2)\in D(U(\gamma))$, for all $n_j\geq0$; (iii ) $\varphi_{n_1,n_2}(x_1,x_2)=U(\gamma)\Phi_{n_1,n_2}(x_1,x_2)$ and $\Psi_{n_1,n_2}(x_1,x_2)=U(-\gamma)\Phi_{n_1,n_2}(x_1,x_2)$, for all $n_j\geq0$. Here $$\Phi_{n_1,n_2}(x_1,x_2)=\frac{1}{\sqrt{n_1!\,n_2!}}\,{A_1^\dagger}^{n_1}{A_2^\dagger}^{n_2}\Phi_{0,0}(x_1,x_2)$$ are the (well known) eigenstates of the hamiltonian $H_0$ in (\ref{31}), which form an o.n. basis for $\Hil$.

Now, to check that $\F_\varphi$ and $\F_\Psi$ are $\G$-quasi bases for a suitable $\G$, we start defining this set as the (finite) linear span of the vectors $\Phi_{n_1,n_2}(x_1,x_2)$, which is dense in $\Hil$. Now\footnote{This is a consequence of the fact that, as we have already seen, the vectors $\varphi_{n_1,n_2}=U(\gamma)\Phi_{n_1,n_2}$ and $\Psi_{n_1,n_2}=U(-\gamma)\Phi_{n_1,n_2}$, for all $n_j\geq0$, are all well defined in $\Hil$.}, since $\G\subseteq D(U(\gamma))\cap D(U(-\gamma))$, we can check that, taken $f,g\in \G$,
$$
\sum_{n_1,n_2}\left<f,\varphi_{n_1,n_2}\right>\left<\Psi_{n_1,n_2},g\right>=\sum_{n_1,n_2}\left<f,U(\gamma)\Phi_{n_1,n_2}\right>\left<U(-\gamma)\Phi_{n_1,n_2},g\right>=
$$
\be
=\sum_{n_1,n_2}\left<U(\gamma)f,\Phi_{n_1,n_2}\right>\left<\Phi_{n_1,n_2},U(-\gamma)g\right>=\left<U(\gamma)f,U(-\gamma)g\right>=\left<f,g\right>.
\label{add2}\en
Hence, $\F_\Psi$ and $\F_\varphi$ are $\G$-quasi bases.

Consequences of the validity of the three assumptions is that we can introduce a new self-adjoint operator,
$\Theta(\gamma)$, such that $(a_j,b_j^\dagger)$ are $\Theta(\gamma)$ conjugate. Moreover, $\Theta(\gamma)$
is positive, and some intertwining relations hold.
{In fact, by considering the relations  between the vectors of $\mathcal F_\varphi$ and $\mathcal F_\Psi$ with the vectors $\Phi_{n_1,n_2}$, we can write:

$$\varphi_{n_1 ,n_2}=U^2(\gamma)\Psi_{n_1 ,n_2}=U(2\gamma)\Psi_{n_1 ,n_2},$$
which, see Section II, suggests to define $\Theta(\gamma)=U(-2\gamma)=U^{-1}(2\gamma)$. Now, using $\Theta(\gamma)$ defined in this way, we can further easily deduce the following
(weak) intertwining relations:
\begin{equation}\label{intertw}
 N_j^\dagger \Theta(\gamma)\varphi_{n_1 ,n_2}=\Theta(\gamma)N_j\varphi_{n_1 ,n_2}, \quad \Theta^{-1}(\gamma)N_j^\dagger\Psi_{n_1 ,n_2}=N_j \Theta^{-1}(\gamma)\Psi_{n_1 ,n_2},
\end{equation}
which, as we can see, are defined respectively on $\F_\varphi$ and $\F_\Psi$, but not of course, on the whole $\Hil$.
}

Moreover, $\varphi_{n_1,n_2}$ are eigenstates of $H=\omega_1b_1a_1+\omega_2b_2a_2+\omega_3\1=\omega_1N_1+\omega_2N_2+\omega_3\1$
with eigenvalues $E_{n_1,n_2}=\omega_1n_1+\omega_2n_2+\omega_3$, while $\Psi_{n_1,n_2}$ are eigenstates of $H^\dagger$ with the same eigenvalues.
Hence $\Theta(\gamma)$ also intertwines (on a suitable domain) between $H$ and $H^\dagger$.
Finally, in terms of the bosonic operators $x_j,p_j$, the deformed hamiltonian $H$ turns out to be the following operator:
\bea
H=x_1^2(\tilde\omega_1-4\gamma^2\tilde\omega_2)+x_2^2(\tilde\omega_2-4\gamma^2\tilde\omega_1)+\tilde\omega_1 p_1^2+\tilde\omega_2 p_2^2-4\sqrt{2}\gamma^2\tilde\omega_2 x_1-4\sqrt{2}\gamma^2\tilde\omega_1 x_2+\nonumber\\
+2\sqrt{2}i\tilde\omega_1 \gamma p_1
+2\sqrt{2}i\tilde\omega_2 \gamma p_2+
4i\tilde\omega_2 \gamma x_1 p_2+4i\tilde\omega_1 \gamma x_2 p_1+\nonumber\\
+(\tilde\omega_3-2\gamma^2 \tilde\omega_1-2\gamma^2 \tilde\omega_2+2\gamma \tilde\omega_1+2\gamma \tilde\omega_2)\1,\label{add3}
\ena
which, apart an additive constant, looks like an asymmetric two-dimensional harmonic oscillator with a manifestly non self-adjoint perturbation, with a linear and a quadratic parts. Therefore we can conclude that the non self-adjoint hamiltonian (\ref{add3}) is just a very complicated way to write a much simpler hamiltonian, whose eigenvalues and eigenvectors can be easily deduced adopting the strategy described in Section II. Also, it is easy to find the eigenvectors of $H^\dagger$ with the same strategy. What it is also interesting is that neither of these two sets are bases for $\Hil$, but they are (both) $\G$-quasi bases.

\subsection{The second model}\label{secIII2}
We now consider the following choice for $X$:

\be
X=\gamma(x_1x_2 +p_1p_2),
\label{320}
\en
where $\gamma\in {\Bbb R}$. This is again quadratic, but compared with our previous choice,
involves also the momentum operators. In (\ref{33}) we assume that $\tilde\omega_1\neq\tilde\omega_2$ since, otherwise, the situation trivializes (see the expression \eqref{316h} for $H$ below),
meaning that the hamiltonian becomes, if $\tilde\omega_1=\tilde\omega_2$, a purely two dimensional, self-adjoint, harmonic oscillator.

As before we introduce $U(\gamma)=e^X$, and using the Baker-Campbell-Hausdorff formula, the operators $a_j$ and $b_j$ are found to be
\be
\left\{
    \begin{array}{ll}
a_1=(C_{\gamma}x_1-S_{\gamma}x_2+i(C_{\gamma} p_1 -S_{\gamma}p_2))/\sqrt{2}=C_{\gamma} A_1-S_{\gamma}A_2,\\
a_2=(C_{\gamma}x_2-S_{\gamma}x_1+i(C_{\gamma} p_2 -S_{\gamma}p_1))/\sqrt{2}=C_{\gamma} A_2-S_{\gamma}A_1,\\
b_1=(C_{\gamma}x_1+S_{\gamma}x_2-i(C_{\gamma} p_1 +S_{\gamma}p_2))/\sqrt{2}=C_{\gamma} A_1^\dagger+S_{\gamma}A_2^\dagger,\\
b_2=(C_{\gamma}x_2+S_{\gamma}x_1-i(C_{\gamma} p_2 +S_{\gamma}p_1))/\sqrt{2}=C_{\gamma} A_2^\dagger+S_{\gamma}A_1^\dagger,\\
       \end{array}
        \right.
\label{321}\en
where $C_{\gamma}:=\cosh{\gamma}$ and $S_{\gamma}:=\sinh{\gamma}$, while the deformed hamiltonian $H$ in (\ref{32}) becomes

\begin{eqnarray}
\label{316h}
H=(\tilde\omega_1 C_{\gamma}^2-\tilde\omega_2 S_{\gamma}^2)x_1^2 +(\tilde\omega_2 C_{\gamma}^2-\tilde\omega_1 S_{\gamma}^2)x_2^2 + (\tilde\omega_1 C_{\gamma}^2-\tilde\omega_2 S_{\gamma}^2)p_1^2+(\tilde\omega_2 C_{\gamma}^2-\tilde\omega_1 S_{\gamma}^2)p_2^2+\\
\nonumber +2i(\tilde\omega_2 - \tilde\omega_1) C_{\gamma} S_{\gamma}(x_1p_2-x_2p_1)+\tilde\omega_3\1.
\end{eqnarray}


Because of the (\ref{321}) it is clear that the vacua of $a_j$, $\varphi_{0,0}(x_1,x_2)$, and of $b_j^\dagger$, $\Psi_{0,0}(x_1,x_2)$, both coincide with the vacuum of the standard bosonic operators $A_j$, $\Phi_{0,0}(x_1,x_2)$:

$$
\varphi_{0,0}(x_1,x_2)=\Psi_{0,0}(x_1,x_2)=\Phi_{0,0}(x_1,x_2).
$$

By introducing the shorthand notation
$$\xi (a,b,c,d)=\sqrt{\frac{(a+b-c-d)!(c+d)!}{a!b!}}, \ \ \ a,b,c,d \in \mathds N,$$
the expressions of $\varphi_{n_1,n_2}$ and $\psi_{m_1,m_2}$  can be written as follows:
$$
\varphi_{n_1,n_2}(x_1,x_2)=\frac{1}{\sqrt{n_1!n_2!}}b_1^{n_1}b_2^{n_2}\Phi_{0,0}(x_1,x_2)= \\$$$$
=\sum_{k=0}^{n_1}\sum_{j=0}^{n_2}\binom{n_1}{k}\binom{n_2}{j}C_{\gamma}^{n_1+j-k}S_{\gamma}^{n_2+k-j}\xi (n_1, n_2, j, k)\Phi_{n_1+n_2-j-k,j+k}(x_1,x_2)=$$
$$
=\binom{n_1+n_2}{n_1}^{\frac{1}{2}}\sum_{k=0}^{n_1}\sum_{j=0}^{n_2}\binom{n_1}{k}\binom{n_2}{j}\binom{n_1+n_2}{j+k}^{-\frac{1}{2}}C_{\gamma}^{n_1+j-k}
S_{\gamma}^{n_2+k-j}\Phi_{n_1+n_2-j-k,j+k}(x_1,x_2),
$$
and
$$
\Psi_{m_1,m_2}(x_1,x_2)=\frac{1}{\sqrt{m_1!m_2!}}(a_1^\dagger)^{m_1}(a_2^\dagger)^{m_2}\Phi_{0,0}(x_1,x_2)=\\$$
$$
=\sum_{i=0}^{m_1}\sum_{l=0}^{m_2}\binom{m_1}{i}\binom{m_2}{l}\bar C_{\gamma}^{m_1+l-i}(-\bar S_{\gamma})^{m_2+i-l}\xi (m_1,m_2,l,i)\Phi_{m_1+m_2-l-i,l+i}(x_1,x_2)=$$$$
=\binom{m_1+m_2}{m_1}^{\frac{1}{2}}\sum_{i=0}^{m_1}\sum_{l=0}^{m_2}\binom{m_1}{i}\binom{m_2}{l}\binom{m_1+m_2}{l+i}^{-\frac{1}{2}}\bar C_{\gamma}^{m_1+l-i}(-\bar S_{\gamma})^{m_2+i-l}\Phi_{m_1+m_2-i-l,i+l}(x_1,x_2).
$$

Hence, as these formulas show, the eigenstates of $H$ and $H^{\dagger}$ (and obviously of $N_j$ and $N_j^{\dagger}$)
are linear combinations of the eigenstates of $H_0$. Using the operator $U(\gamma)$, and repeating the same reasonings as before, they can be written as
$$\varphi_{n_1,n_2}=U(\gamma)\Phi_{n_1 , n_2} \ \ \ \ \ \Psi_{m_1,m_2}=(U^{-1}(\gamma))^{\dagger}\Phi_{m_1 , m_2}=U^{-1}(\gamma)\Phi_{m_1 , m_2}.$$

From the biorthogonality condition we deduce the following, non trivial, summation rule:
$$\left \langle\Psi_{m_1,m_2},\varphi_{n_1,n_2} \right \rangle=$$$$ =\sum_{i=0}^{m_1}\sum_{l=0}^{m_2}\sum_{k=0}^{n_1}\sum_{j=0}^{n_2}\binom{m_1}{i}\binom{m_2}{l}\binom{n_1}{k}\binom{n_2}{j} \xi (n_1, n_2, j, k) \xi (m_1,m_2,l,i) \times$$ $$ \times C_{\gamma}^{m_1+n_1+l-i+j-k}(- 1)^{m_2+i-l}S_{\gamma}^{m_2+n_2+i-l+k-j}\delta_{m_1+m_2,n_1+n_2}\delta_{l+i,j+k}=$$$$
=\delta_{m_1,n_1}\delta_{m_2,n_2}.
$$

To check if the two sets $\mathcal F_{\varphi}$ and $\mathcal F_{\Psi}$ are  bases of $\mathcal L^2 (\mathds R^2)$,
we check, as in the previous section,  whether $\|\varphi_{n_1,n_2}\|$ diverges with $n_j$. For simplicity, we set $n_2 =0$, as in Section \ref{secIII1}. Then
$$
\lVert\varphi_{n_1,0} \rVert^2=$$
$$=\left\langle\frac{1}{\sqrt{n_1!}}\sum_{k=0}^{n_1}\binom{n_1}{k}C_{\gamma}^{n_1-k}S_{\gamma}^k\sqrt{(n_1-k)!k!}\phi_{n_1-k,k},
\frac{1}{\sqrt{n_1!}}\sum_{i=0}^{n_1}\binom{n_1}{i}C_{\gamma}^{n_1-i}S_{\gamma}^i\sqrt{(n_1-i)!i!}\phi_{n_1-i,i}\right\rangle=
$$
$$
=\frac{1}{n_1!}\sum_{k=0}^{n_1}\sum_{i=0}^{n_1}\binom{n_1}{k}\binom{n_1}{i}C_{\gamma}^{2n_1-k-i}S_{\gamma}^{k+i}\sqrt{(n_1-k)!k!}\sqrt{(n_1-i)!i!}\delta_{k,i}=
$$
$$
=\sum_{k=0}^{n_1}\binom{n_1}{k}(|C_{\gamma}|^2)^{n_1-k}(|S_{\gamma}|^2)^k=(|C_{\gamma}|^2+|S_{\gamma}|^2)^{n_1}=\cosh^{n_1}(2\gamma)
$$
Since the hyperbolic cosine of a real number is always greater than one, for each $\gamma\neq0$, the norms of these vectors diverge as $n_1\rightarrow\infty$.
The same results is obviously obtained if we put $n_1=0$  and consider  $n_2\rightarrow\infty$. Similar conclusions can be deduced working with $\Psi_{m_1,m_2}$.
Therefore
the sets of eigenstates of $N_j$ and $N_j^\dagger$ are not biorthonormal bases, although they are still complete in $\mathcal L^2(\mathds R^2)$, for the same reasons discussed in Section \ref{secIII1}.

However, we can check easily, repeating the same arguments as before, that
 $\mathcal F_{\varphi}$ and $\mathcal F_{\Psi}$ are  $\mathcal G$-quasi bases, where $\G$ is the  linear span of the vectors $\Phi_{n_1,n_2}(x_1,x_2)$ as in the first model. 

Due to the validity of the three assumptions $\D$-pb1, $\D$-pb2, $\D$-pbw3, we can consider the self-adjoint operator defined as in the previous section
as $\Theta(\gamma)=U(-2\gamma)$, such that $(a_j,b_j^\dagger)$ are $\Theta(\gamma)$-conjugate. Moreover, $\Theta(\gamma)$
is positive and the same intertwining relations as in \eqref{intertw} hold, and in the same sense.

We conclude that the hamiltonian {introduced in this section}
is not really a {\em new model}, but it is just a sufficiently, but not completely, regular, deformed two dimensional harmonic oscillator. More deformations will be listed in the next section.

\section{Other deformed hamiltonians}\label{secIII3}
This section is devoted to a list of other manifestly non self-adjoint hamiltonians which allow a $\D$-pseudo bosonic treatment,
since they eventually appear to be of the form $H=e^XH_0e^{-X}$, for some suitable $X$, which we don't assume here to be necessarily self-adjoint, and for $H_0$ as in (\ref{31}). Our list is rather concise. We just give the expression of the operator $X$ mapping $H_0$ into $H$, the expression of $H$ itself, and some bibliographic information. On the other hand, what we do not consider here, are all the mathematical details we have discussed in Section \ref{SectIII}, leaving open, for instance, the basis problem for the eigenstates of $H$ and $H^\dagger$, as well as the existence of the sets $\D$ and $\G$.
In other words, some results contained in this section are formal\footnote{This is not true for all the hamiltonians considered in this section, some of which have been treated rigorously in terms of $\D$-PBs already in recent papers.} but, we believe, still interesting in view of possible comparison with the literature: all the hamiltonians we are going to list, can in fact be rewritten, some of them at least formally, some others rigorously, in terms of $\D$-PBs.
Then their eigenvalues and eigenvectors, as well as those of their adjoint, can be deduced quite easily, in principle.

Another aspect which we are not going to consider in this section, and which is relevant for a deeper analysis, is whether the parameters of the transformation $e^X$, see below, should be constrained or not, as it happens Section \ref{secIII1}.

\begin{enumerate}

  \item

 \begin{flalign*} &\begin{cases}
 X=-\alpha(p_1+p_2)+\beta(x_1+x_2), \quad \alpha,\beta \in {\Bbb R},\\
H=\tilde\omega_1(x_1^2+p_1^2)+\tilde\omega_2(x_2^2+p_2^2)+2i\alpha(\tilde\omega_1x_1+\tilde\omega_2x_2)+2i\beta(\tilde\omega_1p_1+\tilde\omega_2p_2)+\\
-\left(\tilde\omega _1+\tilde\omega _2-1\right) \left(\alpha^2+\beta^2\right)\1,\\
\tilde\omega_3=\alpha^2+\beta^2\text{ in \eqref{33}}.
\end{cases} &  \end{flalign*}

This hamiltonian can be found in the literature in \cite{li}, with $\tilde\omega_1=\tilde\omega_2=1/2$.

 \item

 \begin{flalign*} &\begin{cases}
X=-\theta\frac{x_1p_1+p_1x_1}{2}, \quad \theta \in (-\pi/4,\pi/4),\\
H=\tilde\omega_1e^{2i\theta}x_1^2+\tilde\omega_1e^{-2i\theta}p_1^2+\tilde\omega_2x_2^2+\tilde\omega_2p_2^2+\tilde\omega_3\1.
\end{cases} &  \end{flalign*}
This model is the two-dimensional version of the one-dimensional Swanson model as discussed in \cite{prov}, and first introduced in \cite{swa},
with $\tilde\omega_1=\sec(2\theta)/2$.

\item
\begin{flalign*}&\begin{cases}
 X=\gamma(\sqrt{2}(x_1+x_2)+2x_1x_2),\quad \gamma\in {\Bbb R},\\
H=x_1^2(\tilde\omega_1-4\gamma^2\tilde\omega_2)+x_2^2(\tilde\omega_2-4\gamma^2\tilde\omega_1)+\tilde\omega_1 p_1^2+\tilde\omega_2 p_2^2-4\sqrt{2}\gamma^2\tilde\omega_2 x_1-4\sqrt{2}\gamma^2\tilde\omega_1 x_2+\\
+2\sqrt{2}i\tilde\omega_1 \gamma p_1+
+2\sqrt{2}i\tilde\omega_2 \gamma p_2+
4i\tilde\omega_2 \gamma x_1 p_2+4i\tilde\omega_1 \gamma x_2 p_1+\\
+(\tilde\omega_3-2\gamma^2 \tilde\omega_1-2\gamma^2 \tilde\omega_2+2\gamma \tilde\omega_1+2\gamma \tilde\omega_2)\1.
\end{cases}&  \end{flalign*}
This is the hamiltonian introduced in Section \ref{secIII1}
 \item
\begin{flalign*}&\begin{cases}
X=\gamma(x_1x_2 +p_1p_2),\quad \gamma\in {\Bbb R}\\
H=(\tilde\omega_1 C_{\gamma}^2-\tilde\omega_2 S_{\gamma}^2)x_1^2 +(\tilde\omega_2 C_{\gamma}^2-\tilde\omega_1 S_{\gamma}^2)x_2^2 + (\tilde\omega_1 C_{\gamma}^2-\tilde\omega_2 S_{\gamma}^2)p_1^2+(\tilde\omega_2 C_{\gamma}^2-\tilde\omega_1 S_{\gamma}^2)p_2^2+\\
\nonumber +2i(\tilde\omega_2 - \tilde\omega_1) C_{\gamma} S_{\gamma}(x_1p_2-x_2p_1)+\tilde\omega_3\1.\end{cases}&  \end{flalign*}
This is the hamiltonian introduced in Section \ref{secIII2}

 \item
\begin{flalign*}&\begin{cases}
X=\gamma_1(x_1x_2) +\gamma_2(p_1p_2),\quad \gamma_1,\gamma_2\in {\Bbb R},\\
H=x_1^2\left(\tilde\omega _1 C ^2_{\sqrt {\gamma _1 \gamma _2}}-\frac{\gamma _1 \tilde\omega _2 S^2_{\sqrt {\gamma _1 \gamma _2}}}{\gamma _2}\right)+
x_2^2\left(\tilde\omega _2 C ^2_{\sqrt {\gamma _1 \gamma _2}}-\frac{\gamma _1 \tilde\omega _1 S^2_{\sqrt {\gamma _1 \gamma _2}}}{\gamma _2}\right)+\\
+p_1^2\left(\tilde\omega _1 C ^2_{\sqrt {\gamma _1 \gamma _2}}-\frac{\gamma _2 \tilde\omega _2 S^2_{\sqrt {\gamma _1 \gamma _2}}}{\gamma _1}\right)+
p_2^2\left(\tilde\omega _2 C ^2_{\sqrt {\gamma _1 \gamma _2}}-\frac{\gamma _2 \tilde\omega _1 S^2_{\sqrt {\gamma _1 \gamma _2}}}{\gamma _1}\right)+\\
+2ix_1 p_2 S_{\sqrt {\gamma _1 \gamma _2}} C_{\sqrt {\gamma _1 \gamma _2}}\left( \sqrt{\frac{\gamma _1}{\gamma _2}} \tilde\omega _2 - \sqrt{\frac{\gamma _2}{\gamma _1}} \tilde\omega _1 \right)
+2ix_2 p_1 S_{\sqrt {\gamma _1 \gamma _2}} C_{\sqrt {\gamma _1 \gamma _2}}\left( \sqrt{\frac{\gamma _1}{\gamma _2}} \tilde\omega _1 - \sqrt{\frac{\gamma _2}{\gamma _1}} \tilde\omega _2 \right)+\omega_3\1,\\
C_{\sqrt {\gamma _1 \gamma _2}}:=\cosh(\sqrt{\gamma_1 \gamma_2}),\quad S_{\sqrt {\gamma _1 \gamma _2}}:=\sinh(\sqrt{\gamma_1 \gamma_2}).
\end{cases}&  \end{flalign*}

 \item
\begin{flalign*}&\begin{cases}
 X=\gamma(x_1+p_1+x_1 p_1),\quad \gamma\in {\Bbb R},\\
 H=\tilde\omega_1 x_1^2e^{-2i\gamma}+\tilde\omega_2 x_2^2+\tilde\omega_1 p_1^2e^{2i\gamma}+\tilde\omega_2 p_2^2+\\
 +2x_1\tilde\omega_1 e^{i \gamma } \left(-1+e^{i \gamma }\right)+2 p_1\tilde\omega_1e^{i \gamma } \left(-1+e^{i \gamma }\right)+e^{-2 i \gamma } \left(\left(-1+e^{i \gamma }\right)^2 \left(1+e^{2 i \gamma }\right) \tilde\omega _1+e^{2 i \gamma } \tilde\omega _3\right)\1.
\end{cases}&  \end{flalign*}

  \item
\begin{flalign*}&\begin{cases}
 X=\gamma_1(x_1 p_1)+\gamma_2x_1,\quad \gamma_1,\gamma_2\in {\Bbb R},\\
 H=\tilde\omega_1 x_1^2e^{-2i\gamma}+\tilde\omega_2 x_2^2+\tilde\omega_1 p_1^2e^{2i\gamma}+\tilde\omega_2 p_2^2+p_1\frac{2 e^{i \gamma _1} \left(-1+e^{i \gamma _1}\right) \gamma _2  \tilde\omega _1}{\gamma _1}+ \left(\frac{\left(-1+e^{i \gamma _1}\right){}^2 \gamma _2^2 \tilde\omega _1}{\gamma _1^2}+\tilde\omega_3\right)\1.
\end{cases}&  \end{flalign*}

       \item
    \begin{flalign*}&\begin{cases}
 X=\gamma \left(x_1^2-p_1^2+\sqrt{2}x_1\right),\quad \gamma\in {\Bbb R},\\
  H=\tilde\omega_1 x_1^2 (\cos^2{2 \gamma }-\sin^2{2\gamma})+
 \tilde\omega_1 p_1^2 (\cos^2{2 \gamma }-\sin^2{2\gamma})+\tilde\omega_2x_2^2+\tilde\omega_2p_2^2+\\
 +\sqrt{2}\tilde\omega_1 x_1 ( \cos^2{2 \gamma}-\cos{2 \gamma  }-\sin^2{2 \gamma })+i \sqrt{2}\tilde\omega_1 p_1( \cos{2\gamma}\sin{2\gamma}-\sin{2 \gamma })+\\
 +4 i \tilde\omega_1  x_1p_1\cos{2 \gamma } \sin{2 \gamma }+ \\
 +\Bigg(\tilde\omega_1(\frac{1}{2}-\cos{2 \gamma }+\frac{1}{2} \cos^2{2 \gamma }-\frac{1}{2} \sin^2{2 \gamma}+2\cos{2 \gamma } \sin{2 \gamma })+\tilde\omega_3 \Bigg)\1.
  \end{cases}&  \end{flalign*}

       \item
      \begin{flalign*}&\begin{cases}
 X=\gamma \left(x_1^2+p_1^2+x_1\right),\quad \gamma\in {\Bbb R},\\
 H=\tilde\omega_1x_1^2+\tilde\omega_1p_1^2+\tilde\omega_2x_2^2+\tilde\omega_2p_2^2+\tilde\omega_1x_1+(\frac{\tilde\omega_1}{4}+\tilde\omega_3)\1.
   \end{cases}&  \end{flalign*}

       \item
        \begin{flalign*}&\begin{cases}
 X=\gamma \left(x_1^2-p_1^2+2x_1 p_1\right),\quad \gamma\in {\Bbb R},\\
  H= \tilde\omega _1 x_1^2 \left(\cos \left(4 \sqrt{2} \gamma \right)-\frac{i \sin \left(4 \sqrt{2} \gamma \right)}{\sqrt{2}}\right)+
   \tilde\omega _1 p_1^2  \left(\cos \left(4 \sqrt{2} \gamma \right)+\frac{i \sin \left(4 \sqrt{2} \gamma \right)}{\sqrt{2}}\right)+\\
  +\tilde\omega_2x_2^2+\tilde\omega_2p_2^2 +
  x_1 p_1\left( 2i \sqrt{2} \tilde\omega _1 \sin \left(2 \sqrt{2} \gamma \right) \cos \left(2 \sqrt{2} \gamma \right)\right)+\\
  +\left(\tilde\omega_3+ i\sqrt{2} \tilde\omega _1 \sin \left(2 \sqrt{2} \gamma \right) \cos \left(2 \sqrt{2} \gamma \right)\right)\1.
       \end{cases}&  \end{flalign*}

  \item
\begin{flalign*}&\begin{cases}
  X= \gamma_1(x_1+x_2+x_1 x_2)+\gamma_2(p_1+p_2+p_1p_2),\quad \gamma_1,\gamma_2 \in {\Bbb R},\\
 H=x_1^2 \left(C^2_{\sqrt{\gamma _1 \gamma _2}} \tilde\omega _1-\frac{S^2_{\sqrt{\gamma _1 \gamma _2}} \gamma _2 \tilde\omega _2}{\gamma _1}\right)+p_1^2\left(C^2_{\sqrt{\gamma _1 \gamma _2}} \tilde\omega _1-\frac{S^2_{\sqrt{\gamma _1 \gamma _2}} \gamma _1 \tilde\omega _2}{\gamma _2}\right)+ \\
 +x_2^2 \left(C^2_{\sqrt{\gamma _1 \gamma _2}} \tilde\omega _2-\frac{C^2_{\sqrt{\gamma _1 \gamma _2}} \gamma _1 \tilde\omega _1}{\gamma _2}\right)+p_2^2 \left(C^2_{\sqrt{\gamma _1 \gamma _2}} \tilde\omega _2-\frac{S^2_{\sqrt{\gamma _1 \gamma _2}} \gamma _2 \tilde\omega _1}{\gamma _1}\right)+\\
+ x_1\frac{2 \gamma _2 \tilde\omega _2 S^2_{\sqrt{\gamma _1 \gamma _2}}+\gamma _1 \left(\tilde\omega _1 \left(-i \sqrt{\frac{\gamma _2}{\gamma _1}} S_{2 \sqrt{\gamma _1 \gamma _2}}+2 C^2_{\sqrt{\gamma _1 \gamma _2}}-2 C_{\sqrt{\gamma _1 \gamma _2}}\right)-i \sqrt{\frac{\gamma _2}{\gamma _1}} \tilde\omega _2 \left(2 S_{\sqrt{\gamma _1 \gamma _2}}-S_{2 \sqrt{\gamma _1 \gamma _2}}\right)\right)}{\gamma _1}+\\
+p_1\frac{-2 \gamma _1 \tilde\omega _2 S^2_{\sqrt{\gamma _1 \gamma _2}}+\gamma _2 \left(\tilde\omega _1 \left(-i \sqrt{\frac{\gamma _1}{\gamma _2}} S_{2 \sqrt{\gamma _1 \gamma _2}}+2 C^2_{\sqrt{\gamma _1 \gamma _2}}-2 C_{\sqrt{\gamma _1 \gamma _2}}\right)+i \sqrt{\frac{\gamma _1}{\gamma _2}} \tilde\omega _2 \left(2 S_{\sqrt{\gamma _1 \gamma _2}}-S_{2 \sqrt{\gamma _1 \gamma _2}}\right)\right)}{\gamma _2}+ \\
+ x_2\frac{2 \gamma _1 \tilde\omega _1 S^2_{\sqrt{\gamma _1 \gamma _2}}+\gamma _2 \left(\tilde\omega _2 \left(-i \sqrt{\frac{\gamma _1}{\gamma _2}} S_{2 \sqrt{\gamma _1 \gamma _2}}+2 C^2_{\sqrt{\gamma _1 \gamma _2}}-2 C_{\sqrt{\gamma _1 \gamma _2}}\right)-i \sqrt{\frac{\gamma _1}{\gamma _2}} \tilde\omega _1 \left(2 S_{\sqrt{\gamma _1 \gamma _2}}-S_{2 \sqrt{\gamma _1 \gamma _2}}\right)\right)}{\gamma _2}+\\
+p_2\frac{-2 \gamma _2 \tilde\omega _1 S^2_{\sqrt{\gamma _1 \gamma _2}}+\gamma _1 \left(\tilde\omega _2 \left(-i \sqrt{\frac{\gamma _1}{\gamma _2}} S_{2 \sqrt{\gamma _1 \gamma _2}}+2 C^2_{\sqrt{\gamma _1 \gamma _2}}-2 C_{\sqrt{\gamma _1 \gamma _2}}\right)+i \sqrt{\frac{\gamma _2}{\gamma _1}} \tilde\omega _1 \left(2 S_{\sqrt{\gamma _1 \gamma _2}}-S_{2 \sqrt{\gamma _1 \gamma _2}}\right)\right)}{\gamma _1}+ \\
+2 x_2p_1 i C_{\sqrt{\gamma _1 \gamma _2}} S_{\sqrt{\gamma _1 \gamma _2}}\sqrt{\frac{\gamma _1}{\gamma _2}} (\tilde\omega _1-\tilde\omega_2)+2 x_1p_2 i C_{\sqrt{\gamma _1 \gamma _2}} S_{\sqrt{\gamma _1 \gamma _2}} \sqrt{\frac{\gamma _2}{\gamma _1}} (\tilde\omega _2-\tilde\omega_1)+\\
+\1 \Bigg(\tilde\omega_3+C_{\sqrt{\gamma _1 \gamma _2}} S_{\sqrt{\gamma _1 \gamma _2}} \sqrt{\frac{\gamma _1}{\gamma _2}} (\tilde\omega _1-\tilde\omega_2)+ C_{\sqrt{\gamma _1 \gamma _2}} S_{\sqrt{\gamma _1 \gamma _2}} \sqrt{\frac{\gamma _2}{\gamma _1}} (\tilde\omega _2-\tilde\omega_1)+\\
-\frac{2 \left(\tilde\omega _1+\tilde\omega _2\right) S^2_{\frac{1}{2} \sqrt{\gamma _1 \gamma _2}} \left((\gamma _1^2 +\gamma _2^2) \left(C_{\sqrt{\gamma _1 \gamma _2}}+1\right)+2 \gamma _2 \gamma _1 \left(i \left(\sqrt{\frac{\gamma _1}{\gamma _2}}+\sqrt{\frac{\gamma _2}{\gamma _1}}\right)S_{\sqrt{\gamma _1 \gamma _2}}-C_{\sqrt{\gamma _1 \gamma _2}}+1\right)\right)}{\gamma _1 \gamma _2} \Bigg).
\end{cases}&  \end{flalign*}

 \end{enumerate}

\vspace{2mm}

As one can see from this list, in some examples the operator $X$ in invariant under the exchange $(x_1,p_1)\,\leftrightarrow\,(x_2,p_2)$.
This is the case of examples 1, 3, 4, 5 and 11. In the other examples, this exchange produces a different operator $X$ and, recalling that $H_0$ is invariant with respect to this operation, a different $H$, which is still exactly diagonalizable by means of $\D$-PBs.

\vspace{3mm}

For completeness, we conclude this section by listing other hamiltonians which also can be described in pseudo-bosonic terms, and which has already been introduced in the literature along the years. In these cases the transformation is more involved, and will not be given here. These other hamiltonians are:

$$ \left\{
    \begin{array}{ll}
H=H_1+H_2+H_3,\qquad \mbox{where}\\
H_1=\frac{1}{2}(p_1^2+ x_1^2)+\frac{1}{2}( p_2^2+ x_2^2),\\
H_2=\frac{\theta}{2\gamma^2}\,(p_1x_2-p_2x_1)  ,\\
H_3=\frac{i}{\gamma}\left[A(x_1+x_2)+\frac{1}{\gamma^2}\left(p_1\left(B+\theta \frac{A}{2}\right)+p_2\left(B-\theta \frac{A}{2}\right)\right)\right],
       \end{array}
        \right.
$$
$$H=\frac{\nu}{2}\left(p_1^2+x_1^2+p_2^2+x_2^2\right)+i\sqrt{2}\,(p_1+p_2),$$
$$ H=(p_1^2+x_1^2)+(p_2^2+x_2^2+2ix_2)+2\epsilon
x_1x_2,$$
and
$$
H=\frac{1}{2}(p_1^2+x_1^2)+\frac{1}{2}(p_2^2+x_2^2)+i\left[A(x_1+x_2)+B(p_1+p_2)\right].$$

Here $\gamma$, $\nu$, $\theta$, $\epsilon$, $A$ and $B$ are real parameters, see \cite{bagbook}. Also, we recall that a general class of quadratic hamiltonian is discussed, from a different point of view, in \cite{sjo}.

\section{Conclusions}

In this paper we have considered several hamiltonians,
quadratic in two-dimensional position and momentum operators, which
can be analyzed in terms of $\D$-PBs.
This is particularly interesting when the hamiltonian
under analysis is manifestly non self-adjoint.
In this case, we have seen under which conditions on the values of the parameters of the hamiltonian our strategy works properly by checking
the validity of the $\D$-PBs assumptions introduced in Section \ref{sectII}. This was performed  in details in Section \ref{secIII1} and \ref{secIII2}
We have also listed in \ref{secIII3}
several deformed hamiltonians, some of which already introduced elsewhere, which allow, at least formally, a pseudo-bosonic treatment. As for the physical content of the models, we should say that this is under debate, and we would say that these models make physical sense if PT-quantum mechanics, and its relatives, makes sense. Nevertheless, the mathematical aspects of these models appear surely worth of a deeper investigation.

\section*{Acknowledgements}

The authors wish to thank Prof. J. P. Gazeau for useful discussions  during a preliminary stage of their work.
This work was partially supported by the University of Palermo and by G.N.F.M.


\begin{thebibliography}{99}

\bibitem{ben1} C. M. Bender, S. Boettcher, {\em Real Spectra in Non-Hermitian Hamiltonians Having PT-Symmetry}, Phys. Rev. Lett. {\bf 80}, 5243 (1998)

\bibitem{mathpap} E. B. Davies, {\em Pseudospectra, the harmonic oscillator and complex resonances}, Proc. Roy. Soc. London A, {\bf 455},
585-599, (1999); A. Mostafazadeh, {\em Metric Operators for Quasi-Hermitian Hamiltonians and Symmetries of Equivalent Hermitian Hamiltonians}, J. Phys. A: Math. Theo., {\bf 41}, 244017 (2008);  A. Mostafazadeh,    {\em Pseudo-Hermiticity versus PT-Symmetry:  The necessary condition for the reality of the spectrum of a non-Hermitian Hamiltonian}, J. Math. Phys., {\bf 43}, 205-214 (2002); D. Krejcirik and P. Siegl,{\em On the metric operator for the imaginary cubic oscillator}, Phys. Rev. D, {\bf 86}, 121702(R) (2012);  F. Bagarello, A. Inoue, C Trapani, {\em Non-self-adjoint hamiltonians defined by Riesz bases},  J. Math. Phys.,
in press; E. Caliceti, M. Hitrik, S. Graffi, J. Sj\"ostrand, {\em Quadratic $PT$-symmetric operators and similarity with self-adjoint operators},  J. Phys. A: Math. Theor., {\bf 45}, 444007, 2012; S. Albeverio, U. G\"{u}nther, and S. Kuzhel, {\em $J$-Self-adjoint operators with $C$-symmetries: Extension Theory Approach}, J. Phys. A: Math. Theor., {\bf 42}, 105205, 2009.



\bibitem{mathpap2} F. Bagarello, {\em From self to non self-adjoint harmonic oscillators: physical consequences and mathematical pitfalls},
Phys. Rev. A,  {\bf 88}, 032120 (2013); F. Bagarello, A. Fring, {\em A non self-adjoint model on a two dimensional noncommutative space with unbound metric}, Phys. Rev. A, {\bf 88}, doi: 10.1103/PhysRevA.88.042119 (2013)


\bibitem{mathpap3} F. Bagarello, A. Inoue, C Trapani, {\em Weak commutation relations of unbounded operators and applications},  J. Math. Phys., {\bf 52}, 113508,
2011; F. Bagarello, A. Inoue, C Trapani, {\em Weak commutation relations of unbounded operators: nonlinear extensions},  J. Math. Phys., {\bf
53}, 123510, (2012)

\bibitem{bagnewpb} F. Bagarello, {\em More mathematics for pseudo-bosons},  J. Math. Phys., {\bf 54}, 063512 (2013)

\bibitem{bagbook} F. Bagarello, {\em Non-selfadjoint operators in quantum physics: Mathematical aspects},  in {\em Non-selfadjoint operators in quantum physics}, F. Bagarello, J.P. Gazeau, F.H. Szafraniek and M. Znoijl Eds., Wiley, to appear in April 2015

\bibitem{baggar} F. Bagarello and F. Gargano {\em Model pseudofermionic systems: Connections with exceptional points
}, Phys. Rev. A, {\bf 89}, 032113 (2014)


\bibitem{dav} E. B. Davies,  {\em Linear operators and their spectra}, Cambridge University Press, Cambridge (2007)


\bibitem{szego} G. Szeg\"o, {\em Orthogonal Polynomials}, AMS, Providence, (1939)

\bibitem{kolm} A. Kolmogorov and S. Fomine, {\em El\'ements de la th\'eorie des fonctions et de lanalyse fonctionelle}, Mir (1973)

\bibitem{sjo} J. Sj\"ostrand, {\em Parametrices for pseudodifferential operators with multiple characteristics}, Ark. Math., {\bf 12}, 85-130 (1974)



\bibitem{li} Jun-Qing Li, Yan-Gang Miao, Zhao Xue, {\em Algebraic method for pseudo-Hermitian Hamiltonians}, arXiv:1107.4972 [quant-ph]

\bibitem{swa} M. S. Swanson, {\em Transition elements for a
non-Hermitian quadratic Hamiltonian}, J. Math. Phys. {\bf 45}, 585-601
(2004)

\bibitem{prov} J. da Providencia, N. Bebiano, J.P. da Providencia, {\em Non hermitian operators with real spectrum in quantum mechanics}, ELA, 21, 98-109 (2010)




\end{thebibliography}
\end{document}